\documentclass[conference]{IEEEtran}
\makeatletter
\def\ps@headings{%
\def\@oddhead{\mbox{}\scriptsize\rightmark \hfil \thepage}%
\def\@evenhead{\scriptsize\thepage \hfil \leftmark\mbox{}}%
\def\@oddfoot{}%
\def\@evenfoot{}}
\makeatother 
\IEEEoverridecommandlockouts

\usepackage{amsmath}
\usepackage{amssymb}
\usepackage{amsfonts}
\usepackage{graphicx}
\usepackage{epsfig}
\usepackage{subfigure}
\usepackage{psfrag}
\usepackage{cite}
\usepackage{latexsym}
\usepackage{url}
\usepackage{color}
\usepackage{algorithm}
\usepackage{algorithmic}
\usepackage{bbm}

\PassOptionsToPackage{bookmarks={false}}{hyperref}

\begin{document}
\title{Outage Probability Minimization for UAV-Enabled Data Collection with Distributed Beamforming}
\author{\IEEEauthorblockN{Tianxin~Feng\IEEEauthorrefmark{1}, Lifeng~Xie\IEEEauthorrefmark{1}, Jianping~Yao\IEEEauthorrefmark{1}, and Jie~Xu\IEEEauthorrefmark{2}}
\IEEEauthorblockA{\IEEEauthorrefmark{1}School of Information Engineering, Guangdong University of Technology\\
\IEEEauthorrefmark{2}Future Network of Intelligence Institute (FNii) and School of Science and Engineering,\\ The Chinese University of Hong Kong, Shenzhen}
E-mail: ftx.gdut@gmail.com,~lifengxie@mail2.gdut.edu.cn,~yaojp@gdut.edu.cn,~xujie@cuhk.edu.cn

\thanks{J. Yao is the corresponding author.}
}
\setlength\abovedisplayskip{1pt}
\setlength\belowdisplayskip{1pt}
\maketitle

\begin{abstract}
This paper studies an unmanned aerial vehicle (UAV)-enabled wireless sensor network, in which one UAV flies in the sky to collect the data transmitted from a set of sensors via distributed beamforming. We consider the delay-sensitive application scenario, in which the sensors transmit the common/shared messages by using fixed data rates and adaptive transmit powers. Under this setup, we jointly optimize the UAV's trajectory design and the sensors' transmit power allocation, in order to minimize the transmission outage probability, subject to the UAV's flight speed constraints and the sensors' individual average power constraints. However, the formulated outage probability minimization problem is non-convex and thus difficult to be optimally solved in general. To tackle this issue, we first consider the special problem in the ideal case with the UAV's flight speed constraints ignored, for which the well-structured optimal solution is obtained to reveal the fundamental performance upper bound. Next, for the general problem with the UAV's flight speed constraints considered, we propose an efficient algorithm to solve it sub-optimally by using the techniques of convex optimization and approximation. Finally, numerical results show that our proposed design achieves significantly reduced outage probability than other benchmark schemes.
\end{abstract}

\section{Introduction}
Unmanned aerial vehicles (UAVs) or drones are expected to have a lot of applications in beyond-fifth-generation (B5G) and sixth-generation (6G) wireless networks as dedicatedly deployed aerial wireless platforms and cellular-connected aerial users (see, e.g., \cite{ZengAccessing2019,MozaffariBeyond2019,XuUAV2018,XieThroughput2019,XieCommon2020} and the references therein). Among others, there has been an upsurge of interest in using UAVs as aerial data collectors (or fusion centers) to collect data in large-scale wireless sensor networks. Different from the conventional design using on-ground fusion centers for data collection, the UAVs in the sky can exploit the fully controllable mobility in the three-dimensional (3D) space to fly close to sensors for collecting data more efficiently, and can also leverage the strong line-of-sight (LoS) ground-to-air (G2A) channels for increasing the communication quality.

In the literature, there are a handful of prior works studying the UAV-enabled data collection, in which the UAV trajectory is designed for enhancing the system performance (see e.g., \cite{GongFlight2018,LiJoint2019,WangEnergy2019,ZhanEnergy2018,You3D2019,Li2020}). 
For example, the authors in \cite{GongFlight2018} and \cite{LiJoint2019} jointly designed the UAV's flight trajectory and wireless resource allocation/scheduling to minimize the mission completion time, in the scenarios when the sensors are deployed in one-dimensional (1D) and two-dimensional (2D) spaces, respectively. The authors in \cite{WangEnergy2019} and \cite{ZhanEnergy2018} optimized the UAV trajectory and the sensors' transmission/wakeup scheduling, in order to maximize the energy efficiency of the wireless sensor networks while ensuring the collected data amounts from sensors. Furthermore, \cite{You3D2019} exploited the UAV's 3D trajectory optimization for maximizing the minimum average rate for data collection, by considering angle-dependent Rician fading channels. In addition, \cite{Li2020} characterized the fundamental rate limits of UAV-enabled multiple access channels (MAC) for data collection in a simplified scenario with linearly deployed sensors on the ground.
In these prior works, the on-ground devices (or sensors) were assumed to send independent messages to the UAV under different multiple access techniques, and the average data-rate throughput was used as the performance metric by considering the adaptive-rate transmission.

In contrast to communicating independently, distributed beamforming has been recognized as another promising technique to enhance the data rate and energy efficiency in wireless sensor networks (see e.g., \cite{BrownTime2008,MudumbaiDistributed2009,XuWireless2016} and the references therein), in which a large number of sensors are enabled to coordinate in transmitting common or shared messages to a fusion center (the UAV of our interest). By properly controlling the phases, the signals transmitted from different sensors can be coherently combined at the fusion center, thus increasing the communication range and enhancing the energy efficiency via exploiting the distributed beamforming gain. Under this technique, how to jointly design the UAV's trajectory and the sensors' wireless resource allocation for improving the data collection performance is a new problem that has not been investigated in the literature yet.

Motivated by this, this paper focuses on a new UAV-enabled data collection system with distributed beamforming, in which the UAV collects data from  multiple single-antenna sensors via the distributed beamforming. Different from prior works considering the adaptive-rate transmission, we consider the delay-sensitive application scenario (e.g., for real-time video delivery) with adaptive-power but fixed-rate transmission. In this scenario, we aim to minimize the outage probability for data collection by jointly optimizing the UAV's trajectory and the sensors' transmit power allocation over time, subject to the sensors' individual average power constraints and the UAV's flight speed constraints. However, the outage probability minimization problem is non-convex and generally difficult to be optimally solved. To deal with this issue, we first consider the special problem in the ideal case without considering the UAV's flight speed constraints, for which the well-structured optimal solution is obtained to reveal the fundamental performance upper bound. Then, motivated by the obtained trajectory for the above special problem, we propose an efficient approach to obtain a high-quality solution to the general problem with the UAV's flight speed constraints considered, by using techniques from convex optimization and approximation. Finally, numerical results show that our proposed design achieves significantly reduced outage probability as compared with other benchmark schemes.

\section{System Model}
\begin{figure}[!t]
\vspace{-0.1cm}
\setlength{\abovecaptionskip}{-0pt}
\centering
\includegraphics[width=8cm]{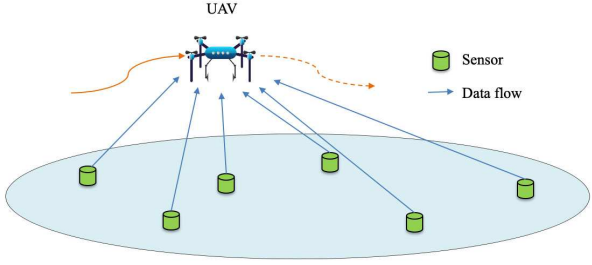}
\caption{Illustration of the UAV-enabled data collection system with distributed beamforming.}	\label{fig:1}
\end{figure}

As shown in Fig. \ref{fig:1}, we consider a UAV-enabled data collection system, in which one single-antenna UAV acts as a mobile date collector to periodically collect data from a set of $\mathcal K\triangleq \{1,\ldots,K\}$ single-antenna sensors on the ground. We assume that all the sensors collaborate as a cluster to transmit common or shared sensing messages towards the UAV with distributed beamforming employed. It is assumed that each sensor $k\in\mathcal K$ is deployed at a fixed location $(x_k,y_k,0)$ on the ground in the 3D Cartesian coordinate system. For notational convenience, let $\boldsymbol{S}_k=(x_k,y_k)$ denote the horizontal location of sensor $k\in\mathcal K$, which is assumed to be known by the UAV {\it a-priori} to facilitate the trajectory design.

We focus on one particular mission period of the UAV with finite duration $T$ in second (s), denoted by $\mathcal{T}\triangleq (0,T]$. The UAV is assumed to fly at a fixed altitude $H$, with the time-varying horizontal location $\boldsymbol{q}(t)=(x(t),y(t))$ for any time instant $t\in\mathcal{T}$. Suppose that $\boldsymbol{q}_{\rm I}$ and $\boldsymbol{q}_{\rm F}$ denote the UAV's initial and final locations, respectively.
Let $V_{\rm max}$ denote the UAV's maximum flying speed. Thus, we
have
\begin{align}
\dot{x}^2(t)+\dot{y}^2(t)\leq V^2_{\rm max}, \forall t\in\mathcal{T},\label{speed}\\
\boldsymbol{q}(0)=\boldsymbol{q}_{\rm I}, \boldsymbol{q}(T)=\boldsymbol{q}_{\rm F},\label{location}
\end{align}
where $\dot{x}(t)$ and $\dot{y}(t)$ denote the first-derivatives of $x(t)$ and $y(t)$ with respect to $t$, respectively.
We also assume that the UAV's mission duration $T$ satisfies $T\geq \|\boldsymbol{q}_{\rm F}-\boldsymbol{q}_{\rm I}\|/V_{\rm max}$, in order for the trajectory from the initial to final locations to be feasible.
Accordingly, the distance between the UAV and sensor $k\in\mathcal K$ at any time instant $t\in\mathcal T$ is given by
\begin{align}
d_{k}(\boldsymbol{q}(t))=\sqrt{\|\boldsymbol{q}(t)-\boldsymbol{S}_k\|^2+H^2}.
\end{align}

As the G2A channels from sensors to UAVs are LoS dominated, we consider a channel model with LoS path loss together with random phases. Consequently, the channel coefficient between the UAV and sensor $k\in\mathcal K$ at any time instant $t\in\mathcal T$ is given by
\begin{align}
{h}_{k}(\boldsymbol{q}(t))=\sqrt{\beta_0d_{k}^{-\alpha}(\boldsymbol{q}(t))}e^{j\psi_{k}(t)},\label{1211}
\end{align}
where $\beta_0$ denotes the channel power gain at the reference distance of $d_0=1$ m, $j=\sqrt{-1}$ denotes the imaginary unit, $\psi_{k}(t)$ denotes the channel phase shift that is uniformly distributed within the interval $[-\pi,\pi]$ \cite{MudumbaiDistributed2009}, and $\alpha\geq2$ denotes the path loss exponent.

In particular, we consider that all the sensors collaborate as a cluster to transmit a common message $s$, which is a circularly symmetric complex Gaussian (CSCG) random variable with zero mean and unit variance (i.e., $s\sim\mathcal{CN}(0,1)$). Such common information can be obtained at different sensors either by their independent sensing (e.g., the common temperature information) or via sharing with each other.\footnote{In order to realize the distributed beamforming, the UAV needs to transmit reference signals over time in order for the sensors to synchronize their transmissions\cite{MudumbaiDistributed2009}.} At any time instant $t\in\mathcal T$, the transmit signal of sensor $k\in\mathcal K$ is $\sqrt{P_{k}(t)}e^{j\varphi_{k}(t)}s$, where $P_k(t)\ge 0$ and $\varphi_{k}(t)\in[-\pi,\pi]$ denote sensor $k$'s transmit power and signal phase, respectively. Suppose that each sensor $k\in\mathcal K$ is subject to a maximum average transmit power $P^{\rm ave}_k$. Therefore, the average transmit power constraint for each sensor $k$ is given by
\begin{align}
\frac{1}{T}\int_{0}^TP_k(t){\rm d}t\leq P^{\rm ave}_k,\forall k\in \mathcal{K}.\label{10202130}
\end{align}
Then, the received signal at the UAV at any time instant $t\in \mathcal{T}$ is given by
\begin{align}
y(t)=\sum_{k=1}^K\sqrt{P_{k}(t)\beta_0d_{k}^{-\alpha}(\boldsymbol{q}(t))}e^{j(\varphi_{k}(t)+\psi_{k}(t))}s+v,\label{201908022328}
\end{align}
where $v$ denotes the additive white gaussian noise (AWGN) at the UAV's information receiver, which is a CSCG random variable with zero mean and variance $\sigma^2$ (i.e., $v\sim\mathcal{CN}(0,\sigma^2)$). In order to achieve the maximum received signal power at the UAV, we design the signal phase as $\varphi_{k}(t)=-\psi_{k}(t), \forall k\in \mathcal{K}, t \in \mathcal{T}$. Thus, the received signal-to-noise ratio (SNR) at the UAV at any time instant $t\in \mathcal{T}$ is given by
\begin{align}
{\tt SNR}(\boldsymbol{q}(t),\{P_k(t)\})\!&=\mathbb{E}_s\!\!\left[\!\left(\!\sum_{k=1}^K\sqrt{P_{k}(t)\beta_0d_{k}^{-\alpha}(\boldsymbol{q}(t))}s\!\right)^2\!\right]\!/\sigma^2\nonumber\\
&=\left(\sum_{k=1}^K\sqrt{P_{k}(t)\beta_0d_{k}^{-\alpha}(\boldsymbol{q}(t))}\right)^2\!/\sigma^2,\label{06051502}
\end{align}
where $\mathbb{E}_s[\cdot]$ denotes the stochastic expectation over the random variable $s$.

In particular, we consider the delay-sensitive application scenario when the sensors use a fixed transmission rate. In order for the UAV to successfully decode the message at any given time instance, the received SNR must be no smaller than a certain threshold $\gamma_{\rm min}$.  In this case, the transmission outage occurs if the received SNR at the UAV falls below $\gamma_{\rm min}$. Therefore, we use the following indicator function to indicate the transmission outage at any time instant $t \in \mathcal{T}$.
\begin{align}
\mathbbm{1}\left({\tt SNR}(\boldsymbol{q}(t),\{P_k(t)\})\right)\nonumber =\left\{
  \begin{array}{ll}
    1,& {\tt SNR}(\boldsymbol{q}(t),\{P_k(t)\}) < \gamma_{\rm min},\\
    0,& {\tt SNR}(\boldsymbol{q}(t),\{P_k(t)\}) \geq \gamma_{\rm min}.
  \end{array}
\right.\label{04181721}
\end{align}
Accordingly, we define the outage probability as the probability that the transmission is in outage over the whole duration $T$, which is expressed as
\begin{align}
O(\{\boldsymbol{q}(t),P_k(t)\})=\frac{1}{T}\int_{0}^T\mathbbm{1}\left({\tt SNR}(\boldsymbol{q}(t),\{P_k(t)\})\right)~{\rm d}t.
\end{align}

Our objective is to minimize the outage probability $O(\{\boldsymbol{q}(t),P_k(t)\})$, by jointly optimizing the UAV's trajectory $\{\boldsymbol{q}(t)\}$ and sensors' power allocation $\{P_k(t)\}$, subject to the UAV's flight speed constraints in (\ref{speed}), the UAV's initial and final locations constraints in (\ref{location}), and the sensors' average transmit power constraints in (\ref{10202130}). Consequently, the outage probability minimization problem of our interest is formulated as
\begin{align}
~~(\mathtt{P1}):\!\!\!\!\!\!\min\limits_{\{\boldsymbol{q}(t),P_k(t)\ge0\}}\!\!\!\!O(\{\boldsymbol{q}(t),P_k(t)\}),
~\mathrm{s.t.}~(\ref{speed}),~(\ref{location}),~\text{and}~(\ref{10202130}).\nonumber
\end{align}
It is worth noting that the objective function of problem $(\mathtt{P1})$ is non-convex and even non-smooth due to the indicator function with coupled variables $\boldsymbol{q}(t)$'s and $P_k(t)$'s. In addition, problem $(\mathtt{P1})$ contains an infinite number of optimization variables over continuous time. As a result, problem $(\mathtt{P1})$ is challenging to be solved optimally.

\section{Proposed Solution to Problem (P1)}\label{s5}
In this section, we first obtain the optimal solution to a relaxed problem of (P1) in the special case with $T\rightarrow\infty$ to gain key engineering insights. Then, based on the optimal solution under the special case, we propose an alternating-optimization-based algorithm to obtain an efficient solution to the original problem $(\mathtt{P1})$ under any finite $T$.

\subsection{Optimal Solution to Relaxed Problem of $(\mathtt{P1})$ with $T\rightarrow\infty$}\label{ss6}
First, we consider the special case that the UAV's flight duration $T$ is sufficiently large (i.e., $T\rightarrow\infty$), such that we can ignore the finite flight time of the UAV from one location to another. As a result, the UAV's flight speed constraints in (\ref{speed}) as well as the initial and final locations constraints in (\ref{location}) can be neglected. Therefore, problem $(\mathtt{P1})$ can be relaxed as
\begin{align}
(\mathtt{P1.1}):~\min\limits_{\{\boldsymbol{q}(t)\},\{P_k(t)\geq0\}}~&O(\{\boldsymbol{q}(t),P_k(t)\}),~~\mathrm{s.t.}~(\ref{10202130}).\nonumber
\end{align}
Though problem $(\mathtt{P1.1})$ is still non-convex, it satisfies the so-called time-sharing condition \cite{YuDual2006}. Therefore, the strong duality holds between problem $(\mathtt{P1.1})$ and its Lagrange dual problem. As a result, we can optimally solve problem $(\mathtt{P1.1})$ by using the Lagrange duality method \cite{boyd2004convex} as follows. 


Let $\mu_k\ge0$ denote the optimal dual variable associated with the $k$-th constraint in (\ref{10202130}). For notational convenience, we define $\boldsymbol\mu\triangleq[\mu_1,\ldots,\mu_K]$. The Lagrangian of problem $(\mathtt{P1.1})$ is given as
\begin{align}
&\tilde{\mathcal{L}}(\{\boldsymbol{q}(t)\},\{P_k(t)\},\boldsymbol\mu)=
\frac{1}{T}\int_{0}^T\mathbbm{1}\left({\tt SNR}(\boldsymbol{q}(t),\{P_k(t)\})\right){\rm{d}}t\nonumber\\
&+\int_{0}^T\sum_{k=1}^K\mu_kP_k(t){\rm{d}}t-T\sum_{k=1}^K\mu_kP_k^{\rm ave}.\label{03010829}
\end{align}
The dual function is
\begin{align}
&\tilde{g}(\boldsymbol\mu)=\min\limits_{\{\boldsymbol{q}(t)\},\{P_k(t)\ge 0\}}\tilde{\mathcal{L}}(\{\boldsymbol{q}(t)\},\{P_k(t)\},\boldsymbol\mu).\label{03010820111}
\end{align}
The dual problem of problem $(\mathtt{P1.1})$ is given by
\begin{align}
(\mathtt{D1.1}):\max\limits_{\{\mu_k \ge 0\}}~~&\tilde{g}(\boldsymbol\mu).
\end{align}
In the following, we solve problem $(\mathtt{P1.1})$ by first obtaining the dual function $\tilde{g}(\boldsymbol\mu)$ and then solving the dual problem ($\mathtt{D1.1}$). First, to obtain $\tilde{g}(\boldsymbol\mu)$, we solve problem (\ref{03010820111}) by solving the following subproblem, in which the index $t$ is dropped  for facilitating the analysis.
\begin{align}
\min\limits_{\boldsymbol{q},\{P_k\geq0\}}~\mathbbm{1}\left({\tt SNR}(\boldsymbol{q},\{P_k\})\right)+\sum_{k=1}^K\mu_kP_k.\label{06061630}
\end{align}
To solve problem (\ref{06061630}), we consider the following two cases when $\mathbbm{1}\left({\tt SNR}(\boldsymbol{q},\{P_k\})\right)$ equals one and zero, respectively.

First, consider that $\mathbbm{1}\left({\tt SNR}(\boldsymbol{q},\{P_k\})\right)=1$. In this case, we have $P_k=0$, and $\boldsymbol q$ can be any arbitrary value. Accordingly, the optimal value for problem (\ref{06061630}) is $1$.

Next, consider that $\mathbbm{1}\left({\tt SNR}(\boldsymbol{q},\{P_k\})\right)=0$. In this case, we solve problem (\ref{06061630}) by first deriving the sensors' power allocation under any given UAV' location $\boldsymbol{q}$ and then search over $\boldsymbol{q}$ via a 2D exhaustive search. Under given $\boldsymbol{q}$ and defining $\rho_k=\sqrt{P_k}, \forall k\in\mathcal{K}$, problem (\ref{06061630}) is reduced as
\begin{align}
\min\limits_{\{\rho_k\ge 0\}}~&\sum_{k=1}^K\mu_k\rho^2_k\label{06061611111}\\
\mathrm{s.t.}~&\sum_{k=1}^{K}\rho_k\sqrt{\beta_0d_{k}^{-\alpha}(\boldsymbol{q})}\geq\sqrt{\gamma_{\rm min}}\sigma\nonumber.
\end{align}
Notice that problem (\ref{06061611111}) is a convex optimization problem. If $\mu_k>0$, then we check the Karush-Kuhn-Tucker (KKT) conditions, and have the optimal solution as
\begin{align}
\rho^{(\boldsymbol\mu,\boldsymbol{q})}_k=\frac{\sqrt{\gamma_{\rm min}\beta_0d_{k}^{-\alpha}(\boldsymbol{q})}\sigma}{\bigg(\sum_{k=1}^K(\beta_0d_{k}^{-\alpha}(\boldsymbol{q})/\mu_k)\bigg)\mu_k}.
\end{align}
If $\mu_k=0$, then problem (\ref{06061611111}) is a linear program, for which the optimal solution of $\{\rho^{(\boldsymbol\mu,\boldsymbol{q})}_k\}$ can be obtained via CVX  \cite{boyd2004convex}. Furthermore, suppose that $P^{(\boldsymbol\mu,\boldsymbol{q})}_k={\rho^{(\boldsymbol\mu,\boldsymbol{q})}_k}^2$. By substituting  $P^{(\boldsymbol\mu,\boldsymbol{q})}_k$ into problem (\ref{06061630}), we can obtain the optimal UAV location $\boldsymbol{q}^{(\boldsymbol\mu)}$ by using the 2D exhaustive search, given as
\begin{align}
\boldsymbol{q}^{(\boldsymbol\mu)}=&\arg \min_{\boldsymbol{q}}\mathbbm{1}\left({\tt SNR}(\boldsymbol{q},\{P_k\})\right)+\sum_{k=1}^K\mu_kP^{(\boldsymbol\mu,\boldsymbol{q})}_k.\nonumber
\end{align}
Accordingly, the obtained power allocation is given by $\{P^{(\boldsymbol\mu,\boldsymbol{q}^{(\boldsymbol\mu)})}_{k}\}$. In this case, the optimal value for problem (\ref{06061630}) is $\sum_{k=1}^K\mu_kP^{(\boldsymbol\mu,\boldsymbol{q}^{(\boldsymbol\mu)})}_k$.

By comparing the corresponding optimal values under $\mathbbm{1}\left({\tt SNR}(\boldsymbol{q},\{P_k\})\right)=1$ and $\mathbbm{1}\left({\tt SNR}(\boldsymbol{q},\{P_k\})\right)=0$, we can obtain the optimal solution to problem (\ref{06061630}) as the one achieving the smaller optimal value. Therefore, the dual function $\tilde{g}(\boldsymbol\mu)$ is obtained.

Next, we solve the dual problem $(\mathtt{D1.1})$ by maximizing the dual function $\tilde{g}(\boldsymbol\mu)$. This is implemented via using subgradient-based methods, such as the ellipsoid method \cite{boyd2008ellipsoid}. We denote the optimal dual solution to $(\mathtt{D1.1})$ as $\boldsymbol{\mu}^{\text{opt}}$.

Finally, with the optimal $\boldsymbol{\mu}^{\text{opt}}$ obtained, it remains to find the optimal primal solution to $(\mathtt{P1.1})$. Notice that under $\boldsymbol{\mu}^{\text{opt}}$, the optimal solution to problem (\ref{06061630}) is non-unique in general.  Suppose that there are $\tilde V$ solutions, denoted by $\{\boldsymbol{q}^{(\boldsymbol\mu^{\text{opt}})}_{\tilde{\nu}}\}$ and $\{P^{(\boldsymbol\mu^{\text{opt}},\boldsymbol{q}^{(\boldsymbol\mu^{\text{opt}})}_{\tilde{\nu}})}_{k}\}$, $\tilde{\nu} = 1,\ldots, \tilde{V}$. In this case, we need to time share among these UAV locations and the corresponding power allocation strategies to construct the primal optimal solution to $\mathtt{(P1.1)}$ as follows.

Let $\tilde{\tau}_{\tilde{\nu}}$ denote the UAV's hovering durations at the location $\boldsymbol{q}^{(\boldsymbol\mu^{\text{opt}})}_{\tilde{\nu}}$, $\tilde{\nu} = 1,\ldots,\tilde{{V}}$. In the following, we solve the following problem to obtain the optimal hovering durations for time sharing.
\begin{subequations}\label{06201515}
\begin{align}
\min\limits_{\{\tilde{\tau}_{\tilde{\nu}}\geq0\}}~& \frac{1}{T}\bigg(T-\sum\limits_{\tilde{\nu}=1}^{\tilde{V}}\tilde{\tau}_{\tilde{\nu}}\bigg)\nonumber\\
\mathrm{s.t.}~&\sum\limits_{\tilde{\nu}=1}^{\tilde{V}}\tilde{\tau}_{\tilde{\nu}}P^{(\boldsymbol\mu^{\text{opt}},\boldsymbol{q}^{(\boldsymbol\mu^{\text{opt}})}_{\tilde{\nu}})}_{k}\leq TP_k^{\rm ave},   \forall k\in \mathcal{K}\\
&\sum\limits_{\tilde{\nu}=1}^{\tilde{V}}\tilde{\tau}_{\tilde{\nu}}\leq T.
\end{align}
\end{subequations}
As problem (\ref{06201515}) is a linear program, the optimal hovering durations $\{\tilde{\tau}^{\text{opt}}_{\tilde{\nu}}\}$ can be obtained by CVX. Therefore, problem ($\mathtt{P1.1}$) is finally solved. Note that at the optimal solution, the UAV hovers at multiple locations over time to collect data from sensors, and the sensors adopt an on-off power allocation, i.e., the sensors are active to send messages with properly designed power allocation when no outage occurs, but inactive with zero transmit power when outage occurs. Also note that if $\mu_k = 0,\forall k\in \mathcal{K}$, then the resulting outage probability is zero (i.e., no outage occurs during the data collection); otherwise, the duration with outage occurring is given by $\tilde{\tau}^{\text{opt}}_0=T-\sum\limits_{\tilde{\nu}=1}^{\tilde{V}}\tilde{\tau}^{\text{opt}}_{\tilde{\nu}}$, with the resulting outage probability being $\tilde{\tau}^{\text{opt}}_0/T$.

\subsection{Proposed Solution to Problem $(\mathtt{P1})$ with Finite $T$}\label{s10}
In this subsection, we consider problem $(\mathtt{P1})$ in the general case with finite $T$. Motivated by the optimal solution to the relaxed problem $(\mathtt{P1.1})$ in the previous subsection, we propose an efficient solution based on the techniques from convex optimization and approximation. Towards this end, we first discretize the whole duration $T$ into a finite number of $N$ time slots denoted by the set $\mathcal N\triangleq\{1,...,N\}$, each with equal duration $\delta=T/N$. Accordingly, problem $\mathtt{(P1)}$ is re-expressed as
\begin{subequations}
\begin{align}
(\mathtt{P1.2})&:\min\limits_{\{\boldsymbol{q}[n]\},\{P_k[n]\geq0\}}~\frac{1}{N}\sum\limits_{n=1}^{N}\mathbbm{1}\left({\tt SNR}(\boldsymbol{q}[n],\{P_k[n]\})\right)\nonumber\\
\mathrm{s.t.}~&\frac{1}{N}\sum\limits_{n=1}^{N} P_k[n]\leq P^{\rm ave}_k, \forall k\in \mathcal{K}\label{08061004}\\
&\|\boldsymbol{q}[n]-\boldsymbol{q}[n-1]\|^2\leq V^2_{\rm max}{\delta}^2, \forall n\in {\mathcal N}\label{08061005}\\
&\boldsymbol{q}[0]=\boldsymbol{q}_I,~\boldsymbol{q}[N]=\boldsymbol{q}_F.\label{IFC}
\end{align}
\end{subequations}
Notice that problem $\mathtt{(P1.2)}$ is still non-convex. To tackle this issue, define $l_n(\boldsymbol{q}[n],\{P_k[n]\})={\tt SNR}(\boldsymbol{q}[n],\{P_k[n]\})-\gamma_{\rm min},\forall n\in \mathcal N$ and $\boldsymbol{l}(\{\boldsymbol{q}[n]\},\{P_k[n]\})=[l_1(\boldsymbol{q}[1],\{P_k[1]\}),\ldots,l_{{N}}(\boldsymbol{q}[{N}],\{P_k[{N}]\})]$. As a result, problem $\mathtt{(P1.2)}$ is equivalently expressed as
\begin{align}
(\mathtt{P1.3}):&~\min\limits_{\{\boldsymbol{q}[n]\},\{P_k[n]\geq0\}}~\frac{1}{N}\|\boldsymbol l(\{\boldsymbol{q}[n]\},\{P_k[n]\})\|_0\nonumber\\
\mathrm{s.t.}~&(\ref{08061004}),~(\ref{08061005}), ~\text{and}~ (\ref{IFC}),\nonumber
\end{align}
where $\|\boldsymbol{x}\|_0$ denotes the zero norm of a vector $\boldsymbol x$ returning the number of non-zero coordinates of $\boldsymbol x$. To handle the zero-norm function in problem $\mathtt{(P1.3)}$, we use $\|\boldsymbol l(\{\boldsymbol{q}[n]\},\{P_k[n]\})\|_1$ to approximate $\|\boldsymbol l(\{\boldsymbol{q}[n]\},\{P_k[n]\})\|_0$ \cite{tropp2006algorithms}. Note that to reduce the outage probability with minimized energy consumption, the received SNR of each time slot should not be larger than $\gamma_{\min}$. Thus, we have the following constraints:
$
{\tt SNR}(\boldsymbol{q}[n],\{P_k[n]\}) \leq \gamma_{\rm min},\forall n\in\mathcal N.
$
By further introducing two sets of auxiliary variables $\{a_k[n]\}$ and $\{A_k[n]\}$, $k\in\mathcal K,n\in\mathcal N$, problem $\mathtt{(P1.3)}$ is approximated as
\begin{subequations}
\begin{align}
(\mathtt{P1.4}):&~\max\limits_{\{\boldsymbol{q}[n]\},\{P_k[n]\geq0\},\{A[n]\},\{a_k[n]\}}~\frac{1}{N}\sum\limits_{n=1}^{N} A[n]/\sigma^2\nonumber\\
\mathrm{s.t.}&~A[n]\le\left(\sum_{k=1}^K a_k[n]\right)^2,\forall n\in\mathcal N \label{12171640}\\
~&a_k[n]\le \sqrt{\frac{P_k[n]\beta_0}{(\|\boldsymbol{q}[n]-\boldsymbol{S}_k\|^2+H^2)^{\alpha/2}}},\forall k\in\mathcal K,n\in\mathcal N \label{eqn:17b}\\
~&A[n]/\sigma^2\le \gamma_{\min},\forall n\in\mathcal N\\
~&(\ref{08061004}),~(\ref{08061005}),~\text{and}~(\ref{IFC}).\nonumber
\end{align}
\end{subequations}
Problem $(\mathtt{P1.4})$ is still non-convex due to the non-convex constraints in (\ref{12171640}) and (\ref{eqn:17b}).

Next, we solve the non-convex problem $(\mathtt{P1.4})$ by optimizing the UAV trajectory and the sensors' power allocation in an alternating manner. First, under any given $\{P_k[n]\ge0\}$, we optimize the UAV trajectory by adopting the successive convex approximation (SCA) technique. In particular, we update the UAV trajectory $\{\boldsymbol{q}[n]\}$ and $\{a_k[n]\}$ in an iterative manner by approximating the non-convex problem into a convex problem. Let $\{\boldsymbol{q}^{(i)}[n]\}$ and $\{a_k^{(i)}[n]\}$ denote the local points at the $i$-th iteration. Under given UAV trajectory $\{\boldsymbol{q}^{(i)}[n]\}$ and $\{{a}^{(i)}_k[n]\}$, since any convex function
is globally lower-bounded by it first-order Taylor expansion at any point, we have the lower
bounds for $\sqrt{\frac{P_k[n]\beta_0}{(\|\boldsymbol{q}[n]-\boldsymbol{S}_k\|^2+H^2)^{\alpha/2}}}$ and $\left(\sum_{k=1}^Ka_k[n]\right)^2$ as follows.
\begin{align}
&\sqrt{\frac{P_k[n]\beta_0}{(\|\boldsymbol{q}[n]-\boldsymbol{S}_k\|^2+H^2)^{\alpha/2}}}\nonumber\\
&\geq\sqrt{P_k\beta_0}\bigg((\|\boldsymbol{q}^{(i)}[n]-\boldsymbol{S}_k\|^2+H^2)^{-\alpha/4}\nonumber\\
&-\frac{\alpha(\|\boldsymbol{q}[n]-\boldsymbol{S}_k\|^2-\|\boldsymbol{q}^{(i)}[n]-\boldsymbol{S}_k\|^2)}{4(\|\boldsymbol{q}^{(i)}[n]-\boldsymbol{S}_k\|^2+H^2)^{\alpha/4+1}}\!\!\bigg)\!\triangleq\!{a^{{\rm low}}_{k(i)}}(\boldsymbol{q}[n]),\label{06191622}\\
&\bigg(\sum_{k=1}^Ka_k[n]\bigg)^2\geq\bigg(\sum_{k=1}^Ka^{(i)}_k[n]\bigg)^2+2\bigg(\sum_{k=1}^Ka^{(i)}_k[n]\bigg)\nonumber\\
&\times \bigg(\sum_{k=1}^Ka_k[n]-\sum_{k=1}^Ka^{(i)}_k[n]\!\!\bigg)\triangleq A^{{\rm low}}_{(i)}(a_k[n]).\label{06191623}
\end{align}
In each iteration $i$ with given local point $\{\boldsymbol{q}^{(i)}[n]\}$ and $\{{a}^{(i)}_k[n]\}$, we replace $\sqrt{\frac{P_k[n]\beta_0}{(\|\boldsymbol{q}[n]-\boldsymbol{S}_k\|^2+H^2)^{\alpha/2}}}$ and $\bigg(\sum_{k=1}^Ka_k[n]\bigg)^2$ as their lower bounds ${a^{{\rm low}}_{k(i)}}(\boldsymbol{q}[n])$ and $A^{{\rm low}}_{(i)}(a_k[n])$, respectively. As a result, the trajectory optimization problem becomes a convex optimization problem, which can be optimally solved by CVX.

Next, under any given UAV trajectory, we optimize the sensors' power allocation by using the SCA technique as well. Similarly as for optimizing the UAV trajectory, we approximate the non-convex terms into convex forms, so as to optimize the UAV trajectory iteratively, for which the details are omitted for brevity. By alternately updating the UAV trajectory and sensors' power allocation, we can obtain a converged solution to problem $\mathtt{(P1.4)}$, which is denoted by $\{\boldsymbol{q}^*[n]\}$ and $\{P_k^*[n]\}$.

Finally, we use an additional step to obtain the sensors' power allocation $\{P_k[n]\}$ for problem $\mathtt{(P1.2)}$ under the obtained UAV trajectory $\{\boldsymbol{q}^*[n]\}$, for which the problem is given as
\begin{align}
(\mathtt{P1.5}):~\min\limits_{\{P_k[n]\geq0\}}~&\frac{1}{N}\sum\limits_{n=1}^{N}\mathbbm{1}\left({\tt SNR}(\boldsymbol{q}^*[n],\{P_k[n]\})\right)\nonumber\\
\mathrm{s.t.}~&\frac{1}{N}\sum\limits_{n=1}^{N}P_k[n]\leq P^{\rm ave}_k, \forall k\in \mathcal{K}.
\end{align}
To solve problem $\mathtt{(P1.5)}$, we sort the time slots based on the SNR $\{{\tt SNR}(\boldsymbol{q}^*[n],\{P^*_k[n]\})\}$, i.e., ${\tt SNR}(\boldsymbol{q}^*[\pi(1)],\{P^*_k[\pi(1)]\}) \ge \cdots\ge {\tt SNR}(\boldsymbol{q}^*[\pi(N)],\{P^*_k[\pi(N)]\})$, with $\pi(\cdot)$ denoting the permutation over $\mathcal N$. Then, we allocate the sensors' transmit power over a subset $\mathcal{N'}$ of time slots with the highest SNR values, i.e., $\mathcal N' = \{\pi(1),\ldots, \pi(N')\}$, where $N'$ is a variable to be determined. To find $N'$ and the corresponding power allocation, we define the following feasibility problem.
\begin{subequations}\label{1112}
\begin{align}
(\mathtt{P1.6}):&{\rm find}~~{\{P_k[n]\geq0\}} \nonumber\\
\mathrm{s.t.}~&{\tt SNR}\left(\boldsymbol{q}^*[\pi(n)],\{P_k[\pi(n)]\}\right) \geq \gamma_{\rm min}, \forall n\in \mathcal{N'}\label{1110}\\
&\frac{1}{N'}\sum\limits_{n=1}^{N'}P_k[\pi(n)]\leq P^{\rm ave}_k, \forall k\in \mathcal{K}.
\end{align}
\end{subequations}
By letting $\rho'_k[n]=\sqrt{P_k}[n]$, problem $(\mathtt{P1.6})$ can be transformed into a convex form and thus be solved optimally via CVX.  By solving problem $(\mathtt{P1.6})$ under given $N'$ together with a bisection search over $N'$, we can find a high-quality solution to problem $(\mathtt{P1.5})$. By combining this together with $\{\boldsymbol{q}^*[n]\}$, an efficient solution of $N'$ and the corresponding power allocation at sensors to problem $(\mathtt{P1})$ is finally obtained.

Note that in order to guarantee the performance of the obtained solution to problem $(\mathtt{P1})$, we need an initial point for iteration. Here, we choose the successive hover-and-fly (SHF) trajectory as the initial point. In SHF trajectory, the UAV flies at the maximum speed from the initial location to successively visit these optimal hovering locations and finally flies to final location. During the flight, we choose the minimum flying path by solving the traveling salesman problem (TSP) (see, e.g., \cite{XuUAV2018}). Suppose that the minimum flying duration among these locations is $T_{\text{fly}}$. If $T<T_{\text{fly}}$, we alternatively consider the direct flight as the initial point, i.e., the UAV flies from initial location to final location directly at a constant speed $\|\boldsymbol{q}_I-\boldsymbol{q}_F\|/T$.

\section{Numerical Results}
\label{s7}
In the simulation, we consider the scenario with $10$ sensors, which are located at $(20,10)$~m, $(30,28)$~m, $(46,0)$~m, $(56,24)$~m, $(94,168)$~m, $(100,200)$~m, $(112,176)$~m, $(162,0)$~m, $(178,40)$~m, and $(200,6)$~m. We set $\beta_0=-30$ dB, $\sigma^2 = -60$ dBm, $K =10$, $\alpha = 2.8$, $V_{\rm max}=40$ m/s, $N = 128$, $H = 50$ m,  $\boldsymbol{q}_{\rm I}=$ ($0,0$) m, $\boldsymbol{q}_{\rm F}= $  ($200,200$)~m, and $\gamma_{\rm min} = 550$.

\begin{figure}[!h]
\centering
\includegraphics[width=8cm]{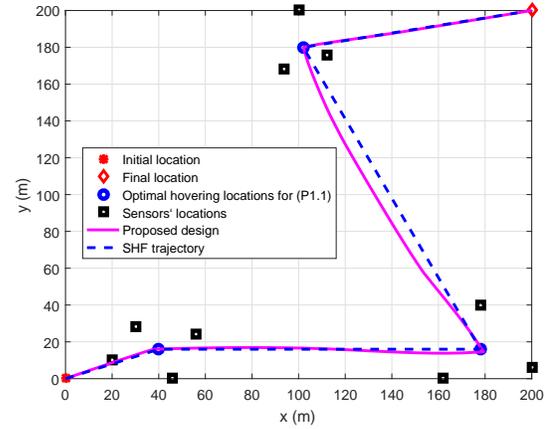}
\caption{System setup and the obtained trajectories with $T=20$~s.}\label{fig:trajectory_sensitive}
\end{figure}

First, Fig. \ref{fig:trajectory_sensitive} shows the system setup and the obtained trajectories with $T=20$~s. It is observed that there are $\tilde{V}=3$ optimal hovering locations for problem $\mathtt{(P1.1)}$.

Next, we compare the performance of our proposed design versus the following three benchmark schemes.
\begin{itemize}
  \item \textbf{Fly-hover-fly trajectory design}: The UAV flies straightly from the initial location to one optimized fixed location $(x^{\text{fix}},y^{\text{fix}},H)$, and then to the final location at the maximum speed. The fixed location $(x^{\text{fix}},y^{\text{fix}},H)$ is obtained via 2D exhaustive search to minimize the outage probability. Under such trajectory, the sensors' power allocation can be obtained by solving problem $(\mathtt{P1.6})$.

  \item \textbf{Power design only}: In this scheme, the UAV flies from the initial location to the final location with a constant flight speed. Under such trajectory, the power allocations at sensors are obtained by solving problem $(\mathtt{P1.6})$.

  \item \textbf{Trajectory design only}: In this scheme, the sensors use the uniform power allocation and accordingly the UAV's trajectory is obtained by iteratively solving problem $\mathtt{(P1.4)}$.
  \end{itemize}

\begin{figure}[!h]
\centering
\includegraphics[width=8cm]{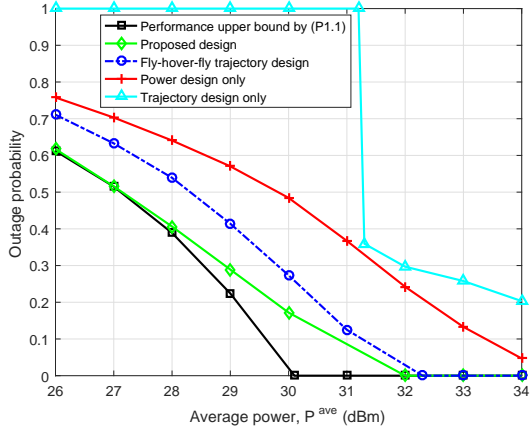}
\caption{Outage probability versus the sensor's maximum average transmit power $P^{\rm ave}$.}\label{fig:power_sensitive}
\end{figure}
Fig. \ref{fig:power_sensitive} shows the outage probability of the system versus the sensor's maximum average power $P^{\rm ave}_k = P^{\rm ave}$, $\forall k\in \mathcal K$, where $T = 20$ s. It is observed that when $P^{\rm ave}$  is less than 31 dBm, the outage probability achieved by the trajectory design only scheme is $1$; while that achieved by other schemes is less than $1$. This shows that power optimization is quite significantly in this case. It is also observed that our proposed design considerably outperforms other benchmark schemes in all regimes of transmit power, by jointly designing the UAV's trajectory and the sensors' power allocation.

\begin{figure}[!h]
\centering
\includegraphics[width=8cm]{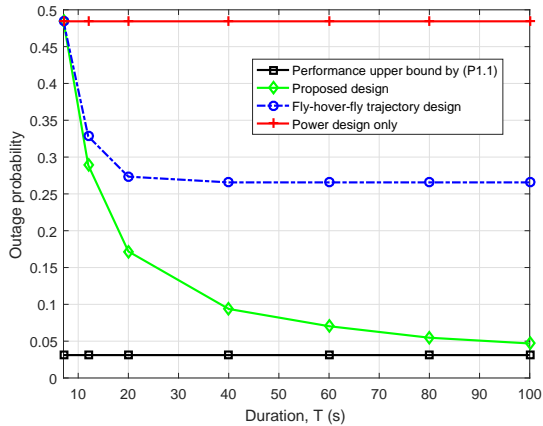}
\caption{Outage probability versus the flight duration $T$.}\label{fig:outage_sensitive}
\end{figure}
Fig. \ref{fig:outage_sensitive} shows the outage probability versus the flight duration $T$, where $P^{\rm ave}_k=30$ dBm, $\forall k\in \mathcal K$. Notice that the trajectory design only scheme always leads to the outage probability of one, and therefore, this scheme is not shown in this figure. It is observed that the proposed design achieves much lower outage probability than the other benchmark schemes, and the performance gain becomes more substantial when $T$ becomes large. Furthermore, with sufficiently large $T$, the proposed design is observed to lead to similar performance as the performance upper bound achieved by problem $\mathtt{(P1.1)}$.

\section{Conclusion}
In this paper, we considered the UAV-enabled data collection from multiple sensors with distributed beamforming. We minimized the transmission outage probability, by jointly optimizing the UAV's trajectory and the sensors' power allocation. To deal with this challenging problem, we first optimally solved the relaxed problem without considering the UAV's flight speed constraints. Next, we used the techniques from convex optimization and approximation to find the sub-optimal solutions to the general problem. Finally, we conducted simulations to show the effectiveness of our proposed design. How to extend our results to other scenarios, e.g., with multiple UAVs and multi-antenna UAVs is an interesting direction worth further investigation.


\footnotesize
\bibliographystyle{IEEEtran}
\bibliography{myreference}

\begin{thebibliography}{10}
\providecommand{\url}[1]{#1}
\csname url@samestyle\endcsname
\providecommand{\newblock}{\relax}
\providecommand{\bibinfo}[2]{#2}
\providecommand{\BIBentrySTDinterwordspacing}{\spaceskip=0pt\relax}
\providecommand{\BIBentryALTinterwordstretchfactor}{4}
\providecommand{\BIBentryALTinterwordspacing}{\spaceskip=\fontdimen2\font plus
\BIBentryALTinterwordstretchfactor\fontdimen3\font minus
  \fontdimen4\font\relax}
\providecommand{\BIBforeignlanguage}[2]{{%
\expandafter\ifx\csname l@#1\endcsname\relax
\typeout{** WARNING: IEEEtran.bst: No hyphenation pattern has been}%
\typeout{** loaded for the language `#1'. Using the pattern for}%
\typeout{** the default language instead.}%
\else
\language=\csname l@#1\endcsname
\fi
#2}}
\providecommand{\BIBdecl}{\relax}
\BIBdecl

\bibitem{ZengAccessing2019}
Y.~{Zeng}, Q.~{Wu}, and R.~{Zhang}, ``Accessing from the sky: A tutorial on
  {UAV} communications for {5G} and beyond,'' \emph{Proc. IEEE}, vol. 107,
  no.~12, pp. 2327--2375, Dec. 2019.

\bibitem{MozaffariBeyond2019}
M.~{Mozaffari}, A.~T.~Z. {Kasgari}, W.~{Saad}, M.~{Bennis}, and M.~{Debbah},
  ``Beyond {5G} with {UAVs}: Foundations of a {3D} wireless cellular network,''
  \emph{IEEE Trans. Wireless Commun.}, vol.~18, no.~1, pp. 357--372, Jan. 2019.

\bibitem{XuUAV2018}
J.~Xu, Y.~Zeng, and R.~Zhang, ``{UAV}-enabled wireless power transfer:
  Trajectory design and energy optimization,'' \emph{IEEE Trans. Wireless
  Commun.}, vol.~17, no.~8, pp. 5092--5106, Aug. 2018.

\bibitem{XieThroughput2019}
L.~{Xie}, J.~{Xu}, and R.~{Zhang}, ``Throughput maximization for {UAV}-enabled
  wireless powered communication networks,'' \emph{IEEE Internet Things J.},
  vol.~6, no.~2, pp. 1690--1703, Apr. 2019.

\bibitem{XieCommon2020}
L.~{Xie}, J.~{Xu}, and Y.~{Zeng}, ``Common throughput maximization for
  {UAV}-enabled interference channel with wireless powered communications,''
  \emph{IEEE Trans. Commun.}, pp. 1--1, 2020.

\bibitem{GongFlight2018}
J.~{Gong}, T.~{Chang}, C.~{Shen}, and X.~{Chen}, ``Flight time minimization of
  {UAV} for data collection over wireless sensor networks,'' \emph{IEEE J. Sel.
  Areas Commun.}, vol.~36, no.~9, pp. 1942--1954, Sep. 2018.

\bibitem{LiJoint2019}
J.~{Li}, H.~{Zhao}, H.~{Wang}, F.~{Gu}, J.~{Wei}, H.~{Yin}, and B.~{Ren},
  ``Joint optimization on trajectory, altitude, velocity and link scheduling
  for minimum mission time in {UAV}-aided data collection,'' \emph{IEEE
  Internet Things J.}, vol.~7, no.~2, pp. 1464--1475, Feb. 2020.

\bibitem{WangEnergy2019}
Z.~{Wang}, R.~{Liu}, Q.~{Liu}, J.~S. {Thompson}, and M.~{Kadoch}, ``Energy
  efficient data collection and device positioning in {UAV}-assisted {IoT},''
  \emph{IEEE Internet Things J.}, vol.~7, no.~2, pp. 1122--1139, Feb. 2020.

\bibitem{ZhanEnergy2018}
C.~{Zhan}, Y.~{Zeng}, and R.~{Zhang}, ``Energy-efficient data collection in
  {UAV} enabled wireless sensor network,'' \emph{IEEE Wireless Commun. Lett.},
  vol.~7, no.~3, pp. 328--331, Jun. 2018.

\bibitem{You3D2019}
C.~{You} and R.~{Zhang}, ``{3D} trajectory optimization in {Rician} fading for
  {UAV}-enabled data harvesting,'' \emph{IEEE Trans. Wireless Commun.},
  vol.~18, no.~6, pp. 3192--3207, Jun. 2019.

\bibitem{Li2020}
P.~{Li} and J.~{Xu}, ``Fundamental rate limits of {UAV}-enabled multiple access
  channel with trajectory optimization,'' \emph{IEEE Trans. Wireless Commun.},
  vol.~19, no.~1, pp. 458--474, Jan. 2020.

\bibitem{BrownTime2008}
D.~R. {Brown III} and H.~V. {Poor}, ``Time-slotted round-trip carrier
  synchronization for distributed beamforming,'' \emph{IEEE Trans. Signal
  Process.}, vol.~56, no.~11, pp. 5630--5643, Nov. 2008.

\bibitem{MudumbaiDistributed2009}
R.~{Mudumbai}, D.~R. {Brown III}, U.~{Madhow}, and H.~V. {Poor}, ``Distributed
  transmit beamforming: challenges and recent progress,'' \emph{IEEE Commun.
  Mag.}, vol.~47, no.~2, pp. 102--110, Feb. 2009.

\bibitem{XuWireless2016}
J.~{Xu}, Z.~{Zhong}, and B.~{Ai}, ``Wireless powered sensor networks:
  Collaborative energy beamforming considering sensing and circuit power
  consumption,'' \emph{IEEE Wireless Commun. Lett.}, vol.~5, no.~4, pp.
  344--347, Aug. 2016.

\bibitem{YuDual2006}
W.~{Yu} and R.~{Lui}, ``Dual methods for nonconvex spectrum optimization of
  multicarrier systems,'' \emph{IEEE Trans. Commun.}, vol.~54, no.~7, pp.
  1310--1322, Jul. 2006.

\bibitem{boyd2004convex}
S.~Boyd and L.~Vandenberghe, \emph{Convex optimization}.\hskip 1em plus 0.5em
  minus 0.4em\relax Cambridge university press, 2004.

\bibitem{boyd2008ellipsoid}
S.~Boyd and C.~Barratt, ``Ellipsoid method,'' \emph{Notes for EE364B, Stanford
  University}, vol. 2008, 2008.

\bibitem{tropp2006algorithms}
J.~A. Tropp, ``Algorithms for simultaneous sparse approximation. {Part II}:
  Convex relaxation,'' \emph{Signal Process.}, vol.~86, no.~3, pp. 589--602,
  Mar. 2006.

\end{thebibliography}
\end{document}